\documentclass[useAMS,usenatbib]{mn2e}

\usepackage{graphicx}
\usepackage{ulem}
\usepackage{float}
\usepackage{psfig}
\usepackage{caption}
\usepackage{subcaption}

\title[Radio emission of ARP\,143]{Multiwavelength study of the radio emission from a tight galaxy pair Arp\,143}
\author[B. Nikiel--Wroczy\'nski et al.]
{
B. Nikiel-Wroczy\'nski\thanks{E-mail:iwan@oa.uj.edu.pl},
M. Jamrozy,
M. Soida, and
M. Urbanik\\
Astronomical Observatory, Jagiellonian
University, ul. Orla 171, Krak\'ow PL 30-244, Poland
}

\begin{document}

\date{Accepted xxxx. Received xxxx; in original form xxxx}

\pagerange{\pageref{firstpage}--\pageref{lastpage}} \pubyear{xxxx}

\maketitle

\label{firstpage}

\begin{abstract}

We present results of the recent low-frequency radio observations of a tight galaxy pair Arp\,143 at 234 and 612\,MHz. 
These data are analysed together with the archive data at 1490, 4860, 8440, and 14940\,MHz. From the analysis of the
radio emission we derive constraints on the age of the radio emitting structures as well as on the properties of their 
magnetic field. We show that the collisional ring of NGC\,2445 hosts strong magnetic fields (reaching 12\,$\mu$G in its
northwestern part) manifesting as a steep--spectrum, nonthermal radiation at radio frequencies. The spectral age of this
structure is higher than estimates derived for the star-forming regions from the H$\alpha$ distribution, suggesting
that the radio emission might have a different origin. The galactic core is of a very young spectral age, suggesting an ongoing 
starburst activity. Additionally we identify a possible ridge of emission between the ring galaxy and its elliptical companion 
NGC\,2444.
\end{abstract}

\maketitle

\section{Introduction}
\label{intro}

 Collisional ring galaxies are scarce, yet very interesting objects. Their most prominent feature is the lack 
of a typical spiral structure, replaced by a narrow, ring--shaped accumulation of gas and stars. Such a peculiar distribution 
of matter is believed to form during collision of a spiral, gas--rich galaxy with a small, early type one (\citealt{LT};
\citealt{TS}). 
Galaxy pair consisting of NGC\,2444 and 2445 is one of the few systems of ring galaxies that have been
included in the Arp's Catalogue, denoted Arp\,143 \citep{arp}. However, it differs significantly from the usual image of
a ring system, as the ring structure in NGC\,2445 is distorted, similar in appearance to a trapezoid with rounded vertices. The
whole visible structure is dominated by local maxima of optical emission that are regions of intensive star formation 
\citep{burridge, tail, multiv}. All of them are relatively young, and the central region is suspected to be undergoing starburst
activity \citep{multiv}. Moreover, the pair is known to be "traling smoke", as the {\rm H}{\sc i} morphology \citep{tail} 
reveals a 150\,kpc long tail of the neutral gas emission extending towards north. All this is suggestive for a collision of 
NGC\,2445 with a compact galaxy -- possibly its companion, NGC\,2444 \citep{multiv,beirao}. This collision disrupted the spiral disk
and resulted in morphological distortions as well as in intensification of the star formation in particular regions of the collisional ring.\\

Not much is known about the radio emission from Arp\,143. In fact, the only work that aimed at revealing the morphology of
the radio-emitting medium of this object was that of \citet{burmil}, who used the Westerbork Synthesis Radio Telescope to study
several interacting galaxies. These authors identified a source of radio emission within NGC\,2445, associating it with one of the 
{\rm H}{\sc ii} regions. However, they excluded the possibility of a purely thermal origin of the emission, as the luminosity was far 
too high to be caused by thermal processes only. Apart from that paper, the only available radio data was that from large surveys (e.g.
\citealt{sulentic} or \citealt{davis}). \citet{colrings} mention a study of Arp\,143 (as well as of ten other ring systems) made with the
Very Large Array (VLA) at the frequency of 8440\,MHz, but no detailed information is provided. Using Arp\,10 as an example, these Authors 
suggest that the spectral index of the emission from ring galaxies might steepen inwards from the ring. Such phenomenon could be a result of
a change from thermal to non-thermal emission and/or of ageing of the highest energy electrons, which leads to their absence in the inner 
region of the ring. However, almost nothing is known about the magnetic field and its properties. Also, cross--identifications between 
radio-emitting structures and their counterparts in other domains of the electromagnetic spectrum have not yet been made.\\

In this paper we present results from our recent observing project at the Giant Metrewave Radio Telescope (GMRT), in which Arp\,143 has
been studied at 234 and 612\,MHz. The new observations are analysed together with the previously unpublished, archive VLA data at 1490, 
4860, 8440, and 14940\,MHz to produce high resolution maps of the radio emission and to study the magnetic field of this galaxy pair.

\begin{figure*}
\resizebox{\hsize}{!}{\includegraphics{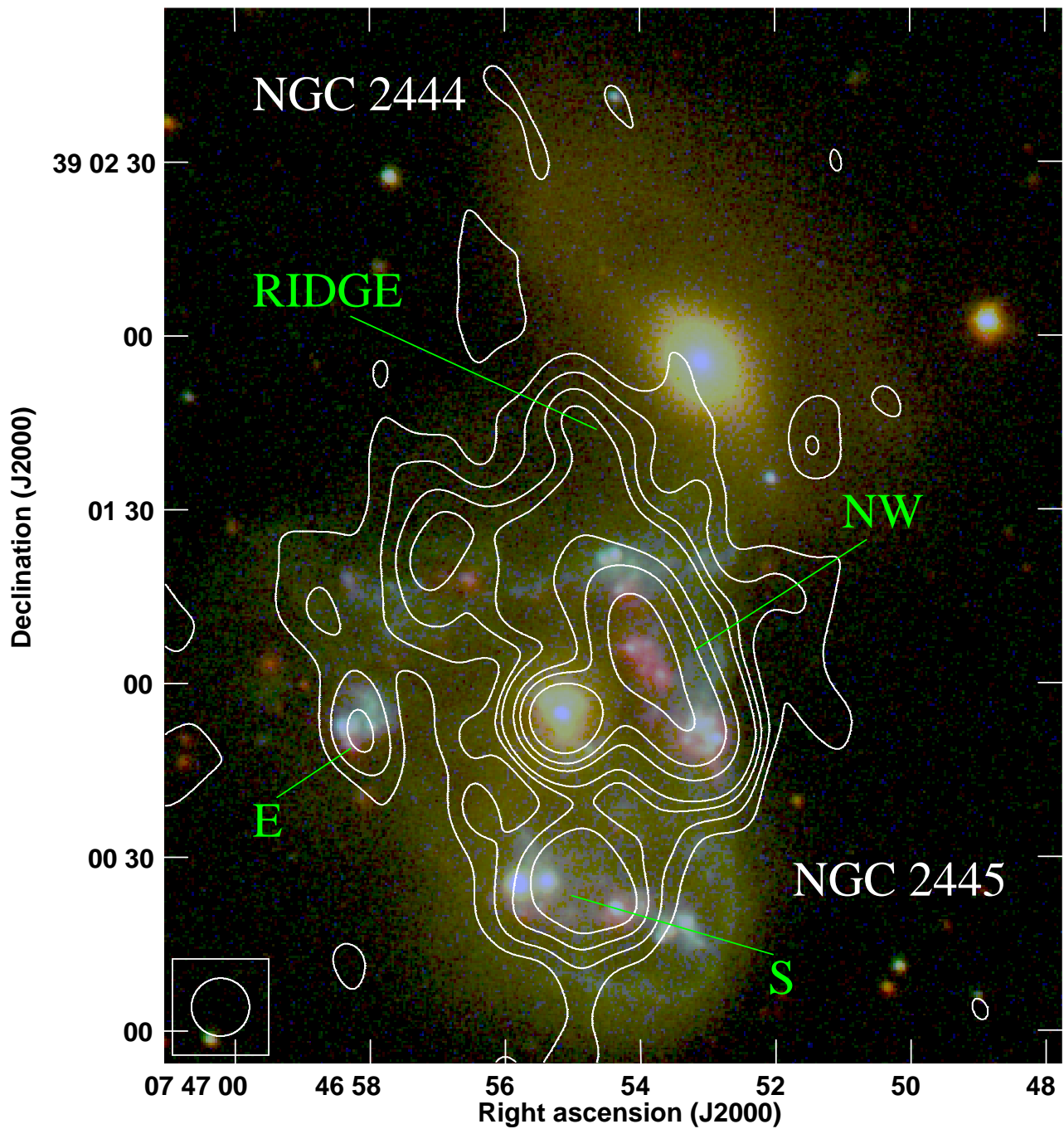}}
\caption{
GMRT map of the TP emission from Arp\,143 at 612\,MHz overlaid upon an RGB image,
with locations of the star-forming regions and intergalactic ridge indicated.
The contour levels are $3,5,10,20,30,50,100 \times$ 0.1\,mJy/beam
(r.m.s. noise level). The angular resolution is 10\,arcsec.   
The beam is represented by a circle in the lower left corner of the image.
Details of the maps used to produce this RGB composite can be found in the text
(Sect.~\ref{result}).
}
\label{RGB}
\end{figure*}

\section{Observations and data reduction}
\label{observ}

\subsection{GMRT data}
\label{obsgmrt}

The GMRT near Pune, India was used to observe Arp\,143 at 234 and 612\,MHz. The observations were carried out in a dual 
frequency mode as a part of our project ``Magnetic field and galaxy interactions -- from loose
groups to mergers'' (project code 23\_025). The observations were carried out in February, 2013.
The total observing time was 8\,h and the bandwidths were 16 and 32\,MHz at 234 and 612\,MHz, respectively. The (u,v) data
were reduced using the Astronomical Image Processing System (\textsc{aips}), including calibration and RFI--flagging. After 
obtaining initial images at both frequencies, they were processed by a self-calibration pipeline in order to correct
the phase information. The final images have been (u,v)-tapered to obtain circular beams and then corrected for the primary 
beam shape. Two images have been made: one with resolution of 16\,arcsec at 234\,MHz, and another at 612\,MHz, with 
resolution of 10\,arcsec. Original beam sizes and noise levels of these maps can be found in Table~\ref{beamsizes}.

\begin{table*}
\caption{\label{beamsizes}Basic information on the observational datasets used in this study}
\begin{center}
\begin{tabular}{rrrrrrrl}
\hline
\hline
Freq. [MHz] & Telescope & Proj. code & Date & TOS [h] & Org. beam [arcsec] & Fin. beam [arcsec] & 
Noise [mJy\slash beam]\\
\hline
  234 & GMRT  & 23\_025 & 10.02.2013 & 8.0 &  14x11  &      16 & 0.4 \\
  612 & GMRT  & 23\_025 & 10.02.2013 & 8.0 & 7.5x5   &      10 & 0.1 \\
 1490 & VLA B & AD 182  & 05.09.1986 & 1.5 &   4x4   &      10 & 0.08\\
 4860 & VLA C & AD 182  & 14.11.1986 & 2.7 &   4x4   &      10 & 0.02\\
 8440 & VLA D & AA 146  & 25.08.1992 & 0.5 &   9x7   &      10 & 0.06\\
14940 & VLA C & AJ 105  & 29.05.1984 & 2.8 & 1.3x1.1 & 1.3x1.1 & 0.08\\
\hline
\end{tabular}
\end{center}
\end{table*}

\subsection{Archive VLA data}
\label{obsvla}

In order to construct the radio continuum spectrum, and to derive the spectral age as well as to provide the magnetic field 
strength estimates for Arp\,143, we searched through the archive of the VLA of the National Radio  Astronomy Observatory (NRAO)
\footnote{NRAO is a facility of National  Science
Foundation operated under cooperative agreement by Associated Universities,  Inc.}. We have obtained archived
data at 1490, 4860, 8440, and 14940\,MHz. Details on these observations can be found in Table~\ref{beamsizes}. All 
these sets were calibrated, flagged and imaged using the \textsc{aips}. The 4860\,MHz data allowed to run self-calibration.
Except for the highest frequency, all the data sets were (u,v)--tapered to obtain a circular beam of 10\arcsec.
The 14940\,MHz image has not been tapered, as it was intended to be used only to derive the size of the galactic core of 
NGC\,2445 and its flux. Additionally, the size of the primary beam at this frequency is nearly the same as the size of NGC\,2445, 
resulting in non-reliable flux values besides the very center of this galaxy.

To include the calibration uncertainties, we have assumed a 5 per cent error for each integrated flux value for all the radio maps
except the 234\,MHz map, for which we adopt 8 per cent error.

\section{Results}
\label{result}

Fig.~\ref{RGB} shows a composite RGB image built from \textit{u, g, r} bands of the Sloan Digital Sky Survey (SDSS) data. An additional, 
H$\alpha$ component (map from \citealt{romano}, taken from the NASA Extragalactic Database) was added to the red channel. Before merging,
the H$\alpha$ map was rescaled to have signal values significantly higher than the median value of the \textit{r} band emission in order to make the
regions of molecular emission easily distinguishable. The colour image was then overlaid with the contours of
radio emission at 612\,MHz. Designations of different structures described in this paper -- namely southern (S), northwestern (NW) and eastern 
(E) star-forming regions, intergalactic ridge and both galaxies -- have also been marked. Measured fluxes of these structures are presented 
in Table~\ref{fluxess}.

\begin{table*}
\caption{Flux densities (with errors) obtained for the selected regions of radio emission. All values given in mJy. ND means no detection.}
\begin{center}
\begin{tabular}[]{lccccccc}
\hline
\hline
Region & 234 MHz & 612 MHz & 1420 MHz & 4860 MHz & 8440 MHz & 14960 MHz \\
\hline
Core  & 15.58 $\pm$ 1.40 & 11.14 $\pm$ 0.57  & 6.92 $\pm$ 0.36 &  3.59 $\pm$ 0.26 &  1.86 $\pm$  0.11 & 0.81 $\pm$ 0.09 \\
NW    & 30.97 $\pm$ 2.61 & 20.07 $\pm$ 1.01  & 9.33 $\pm$ 0.49 &  3.24 $\pm$ 0.16 &  1.39 $\pm$  0.12 &       ND        \\
S     &        ND        &  3.57 $\pm$ 0.25  &       ND        &  0.57 $\pm$ 0.04 &        ND         &       ND        \\
E     &        ND        &  0.66 $\pm$ 0.11  &       ND        &  0.08 $\pm$ 0.02 &        ND         &       ND        \\
Ridge &  3.30 $\pm$ 0.69 &  1.28 $\pm$ 0.10  &       ND        &  0.16 $\pm$ 0.03 &        ND         &       ND        \\
\hline
\end{tabular}
\end{center}
\label{fluxess}
\end{table*}

\begin{figure*}
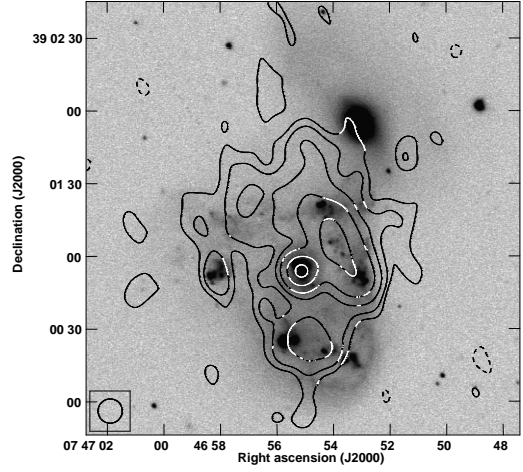
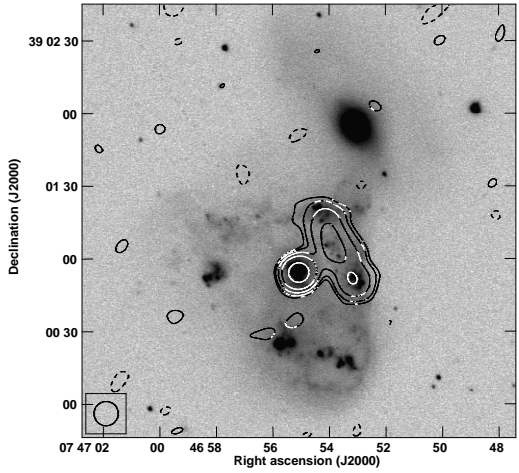
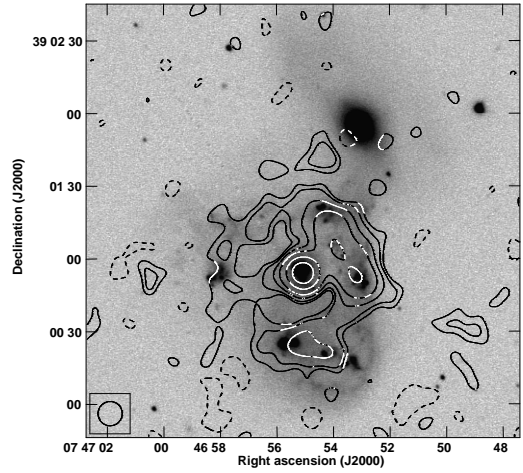
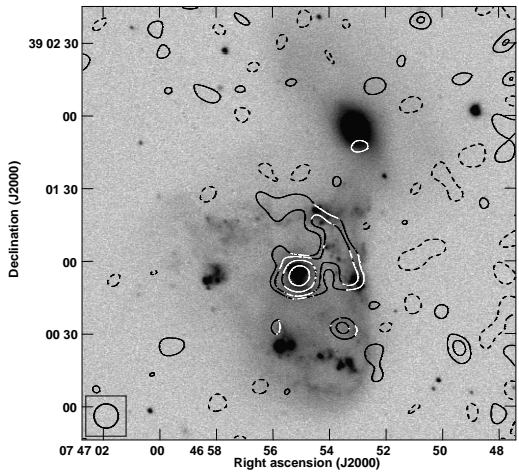
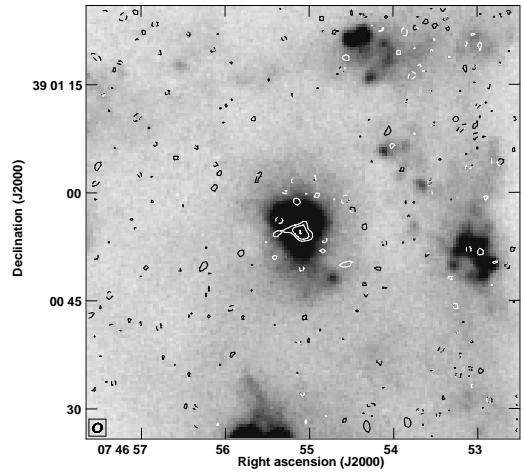

    \begin{subfigure}[p]{0.4\textwidth}
    \includegraphics[width=\textwidth]{pbandN.ps}
    \captionsetup{width=\textwidth}
    \caption{
      GMRT map at 234\,MHz.  
      The r.m.s. noise level is\\ 0.4\,mJy/beam.
      The angular resolution is 16\,arcsec.   
    }
    \label{pband}
    \end{subfigure}%
    \begin{subfigure}[p]{0.4\textwidth}
    \includegraphics[width=\textwidth]{gbandN.ps}
    \captionsetup{width=0.95\textwidth}
    \caption{
      GMRT map at 612\,MHz.  
      The r.m.s. noise level is 0.1\,mJy/beam.
      The angular resolution is 10\,arcsec.   
    }
    \label{gband}
    \end{subfigure}
    
    \begin{subfigure}[p]{0.4\textwidth}
    \includegraphics[width=\textwidth]{lbandN.ps}
    \captionsetup{width=\textwidth}
    \caption{
      VLA map at 1490\,MHz.  
      The r.m.s. noise level is\\ 0.08\,mJy/beam.
      The angular resolution is 10\,arcsec.   
    }
    \label{lband}
    \end{subfigure}%
    \begin{subfigure}[p]{0.4\textwidth}
    \includegraphics[width=\textwidth]{cbandN.ps}
    \captionsetup{width=0.95\textwidth}
    \caption{
      VLA map at 4860\,MHz.  
      The r.m.s. noise level is 0.02\,mJy/beam.
      The angular resolution is 10\,arcsec.   
    }
      \label{cband}
     \end{subfigure}

    \begin{subfigure}[p]{0.4\textwidth}
    \includegraphics[width=\textwidth]{xbandN.ps}
    \captionsetup{width=\textwidth}
    \caption{
      VLA map at 8440\,MHz.  
      The r.m.s. noise level is\\ 0.06\,mJy/beam.
      The angular resolution is 10\,arcsec.   
    }
      \label{xband}
      \end{subfigure}%
    \begin{subfigure}[p]{0.4\textwidth}
    \includegraphics[width=\textwidth]{ubandN.ps}
    \captionsetup{width=0.95\textwidth}
    \caption{
      VLA map of the core at 14940\,MHz.  
      The r.m.s. noise level is 0.08\,mJy/beam.
      The angular resolution is 1.28x1.13\,arcsec.   
    }
	\label{uband}
    \end{subfigure}
    \caption{
	  Maps of the TP emission from Arp\,143 at various frequencies overlaid upon an SDSS {\it g}-band image.
	  The contour levels are $-3$ (dashed), $3,5,10,20,50,100\times$ r.m.s. noise level. 
	  The beam is represented by an ellipse in the lower left corner of the image.
	}
\end{figure*}

The lowest frequency used in our study is 234\,MHz and the map at this frequency is presented in Fig.~\ref{pband}. It has the lowest 
resolution (circular beam of 16 arcseconds in diameter). The emission concentrates in the central-northern region of NGC\,2445 and extends 
northwest, vanishing in the outer regions of NGC\,2444 (which is otherwise a non-radio-emitting galaxy). The central part consists of the core 
and the NW region. Because of the beam size, the emission coming from these two entities is not separated. The radio emission coincides almost 
perfectly with the optically--emitting material in the northern part of the ring, but diminishes abruptly about 10 arcseconds south from the core
of NGC\,2445. Only an isolated patch of emission coincident with the southern star-forming region was detected. This is caused by the lower sensitivity
to faint, extended structures at 234\,MHz compared to 612\,MHz.

The map at 612\,MHz (Fig.~\ref{gband}) has a modest resolution of 10 arcsec and a long integration time, resulting in a deep and detailed radio map. 
Despite higher resolution and frequency, this map shows more extended emission than that at 234\,MHz. Apart from the core and the NW region, two
other star-forming areas can be easily identified. Both in the eastern and the southern part of the ring, local maxima of the visible light 
distribution are coincident with peaks of the radio emission; the eastern structure is the weakest one. The northern extension that forms a 
ridge--like structure between NGC\,2444 and NGC\,2445 is also visible; it extends even further than that at 234\,MHz, reaching NGC\,2444.

The map at 1490\,MHz (made from archive data) is presented in Fig.~\ref{lband}. Short integration time and lack of the shortest baselines (due to the
wide B-configuration of the VLA) resulted in the absence of most of the extended emission as well as in higher noise level. Only the core and the NW region
are visible. There is no trace of emission from other parts of the collisional ring, nor from the aforementioned intergalactic ridge.
It should be noted that the archive VLA data used in this study are of worse (u,v) coverage than the GMRT ones, resulting in lower detectability of the weak, extended
emission. This mostly applies to the 1490 and 8440\,MHz data, for which the (u,v) plane sampling is poor due to the integration times lower than
2 hours.

\begin{figure}
\resizebox{\hsize}{!}{\includegraphics{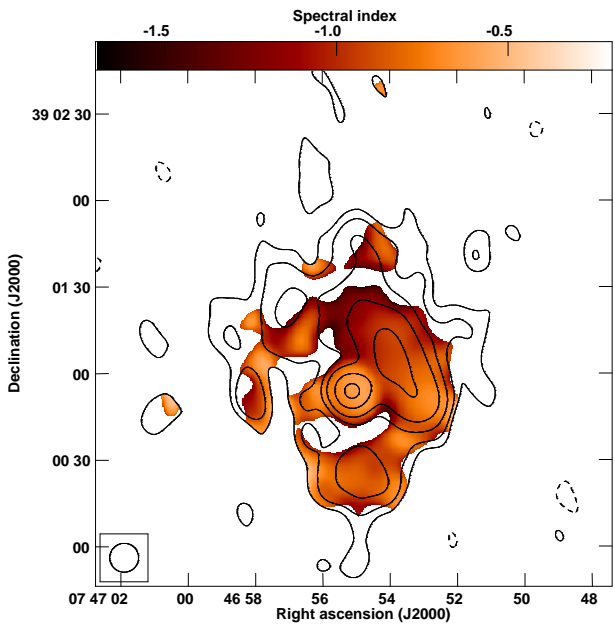}}
\caption{
 Map of the spectral index calculated between 612 and 4860\,MHz
 with contours of the radio emission at 612\,MHz overlaid.
 The contour levels are $-3$ (dashed), $3,5,10,20,30,50\times$ 0.02\,mJy/beam
 (r.m.s. noise level). The angular resolution is 10\,arcsec. 
}
\label{spixGC}
\end{figure}

Considerably better (u,v) coverage and lack of strong RFI allowed to obtain a detailed map of the extended emission at 4860\,MHz (Fig.~\ref{cband}). 
Similarly to the 612\,MHz map, not only the emission from the NW star formation region 
and the core can be seen, but the whole collisional ring is emitting at this frequency. From its structure one can distinguish local maximum corresponding
to the southern star forming region. Surprisingly, the eastern star forming region is indistinguishable from the extended emission. A patch of emission
north from the ring, spatially coincident with the ridge detected both at 234 at 612\,MHz is easily visible. It should be noted here that, unlike at 612\,MHz,
the inner part of the galaxy -- an area between the ring and the core -- is not visible in emission at 4860\,MHz; this effect is more pronounced in the 
spectral index map (Fig.~\ref{spixGC}), indicating a significantly steeper spectrum there.

At 8440\,MHz the NW region is still visible, albeit significantly less prominent than at lower frequencies. Most of the extended emission has not been detected,
because of the very short integration time of these archival data. Despite that, there is some minor extension of the NW region into the northern part of the 
collisional ring, but it barely exceeds the 3\,$\sigma$ level (Fig.~\ref{xband}). Throughout the external parts of the image several isolated patches of 
noise--level radio emission, boosted by the primary beam correction are visible.

The primary beam of the observations at the highest frequency (14940\,MHz) is very small, therefore only the very central part of NGC\,2445 is visible.
Only the core emerges from the noise (Fig.~\ref{uband}). With the high resolution of this map, we can estimate the angular size of the radio-emitting region,
which is approximately 3x2\,arcseconds (at the 3\,$\sigma$ level). This (assuming the distance estimate of $54.9 \pm 3.8$\,Mpc from \citealt{devac}) yields 
the linear size of approximately 0.8 by 0.5\,kpc. Similarly to the 8440\,MHz image (Fig.~\ref{xband}), loose patches of emission resulting from the primary 
beam correction are visible in the outer parts of the image.

Large primary beams in our low-frequency observations allowed us to search for the continuum counterpart of the giant {\rm H}{\sc i} tail extending north 
from the ring galaxy \citep{tail}. However, nothing was detected above the noise level in all the maps. Some remarks on the upper limit of the magnetic 
field strength in this area can be found in Sect.~\ref{magfield}.

\section{Discussion} 
\label{discus}

\subsection{Spectral index}
\label{spixtext}

The spectral index map has been calculated between 612 and 4860\,MHz and has a resolution of 10 arcsec. The input maps have been clipped 
at the 5$\sigma$ level to ensure that the noise fluctuations would not be taken into account. The choice of this two maps -- instead of
the ones at the lowest and highest frequencies -- was dictated by their resolution and quality. The 234\,MHz map (Fig.~\ref{pband})
has a rather low resolution of 16 arcsec, which does not allow to clearly distinguish different emitting regions. Moreover, despite large
beamsize, it does not show larger extent of emission than the 612\,MHz map. The 8440\,MHz map (Fig.~\ref{xband}) has a modest resolution
of 10 arcsec, but it suffers from the short integration time, resulting in significant losses of the extended flux. This is not the case of 
the 4860\,MHz map (Fig.~\ref{cband}), which clearly shows the ring structure as well as a patch spatially connected to the intergalactic 
ridge. Therefore, the spectral index map -- as well as all spectral index values used and presented in this study -- have been calculated 
between the best-quality 612 and 4860\,MHz maps. Throughout the paper we are using the $S_{\nu} \propto \nu^{-\alpha}$ definition of the 
spectral index $\alpha$.

The emission from the core of NGC\,2445 has a steeper spectrum (a mean spectral index of $\approx 0.64 \pm 0.08$) than expected if it were
a purely thermal source. This indicates that it has a synchrotron origin, as we would expect the thermal emission to manifest with a 
significantly flatter spectrum, which is not the case here (see \citealt{pacholczyk} for details). 

The spectrum of the star forming regions is steeper, with the mean spectral indices typical for an ageing population of relativistic electrons.
The values for the particular regions are $\alpha \approx 0.81 \pm 0.07$ for the northwestern, $\alpha \approx 1.03 \pm 0.19$ for the eastern
and $\alpha \approx 0.89 \pm 0.06$ for the southern region. This means that the synchrotron emission dominates everywhere over the thermal component,
even at the higher of these frequencies. 
This is not unusual; \citet{niklas28} have shown that the median thermal fraction at 1\,GHz is $ 0.08 \pm 0.01$. The median non-thermal 
spectral index given by these authors is $0.83 \pm 0.02$, very close to the values derived for the star-forming regions in NGC\,2445.
This means domination of the high--energy, relativistic electrons supplied by the supernovae. \citet{colrings} suggest that
this process of electron supply would start some $10^6$ years after the collision. The estimated age of the density wave that gave birth to 
the NW region is $\approx$ 85\,Myr \citep{beirao}, indicating that such scenario is possible. The inner part of the ring has not
been detected at 4860\,MHz, indicating a very steep spectrum there. Assuming the 3-r.m.s. level as a constraint on the emission at 4860\,MHz, we have
estimated the spectral index to be $> 1.8$ between 612 and 4860\,MHz. Steepening of the spectral index inwards from the ring was mentioned
by \citet{colrings}, who provided two possible explanations for such phenomenon: change from the mostly thermal to non-thermal radiation and/or
ageing of the synchrotron electrons. Our findings suggest that in case of Arp\,143 the second scenario is more probable.

The intergalactic ridge has a steep spectrum -- characteristic for an ageing population of electrons -- with a mean spectral index 
$\alpha \approx 1.01 \pm 0.12$. This is consistent with its identification as an intergalactic structure presented in Sect.~\ref{result}. Also, it is
steeper than that of the star-forming regions NW and S, suggesting a higher spectral age, as it is far away from the star-forming
areas.

\subsection{Magnetic field}
\label{magfield}

Both the strength of the magnetic field $B_{\mathrm{TOT}}$ and its energy density $E_{\mathrm{B}}$ in the selected areas were calculated
assuming the energy equipartition between the cosmic rays and the magnetic field, following the formulae presented in \citet {bfeld}.
We applied the \textsc{bfeld} code, which uses following parameters: total pathlength through the source $D$, proton--to--electron energy density ratio
$K_0$, spectral index $\alpha$, and the mean synchrotron surface brightness of the chosen region to estimate its total field strength as well as the 
magnetic energy density. The mean brightness has been obtained from the region's integrated flux density $S$ by dividing it by the square root of the 
number of the beams in the integration area. $K_0$ was fixed as 100; such a value is suggested also for starburst galaxies \citep{beckstarburst}. Values
of all parameters used, together with the results of the estimation are presented in Table~\ref{params}. The spectral index was calculated between the 
maps at 612 and 4860\,MHz, which have the best quality. 
According to \citet{niklas28}, thermal fraction at 1490\,MHz for majority of galaxies does not exceed 10 per cent; even at 10000\,MHz its mean value is about 
25 per cent. This consistently yields thermal fraction at 4860\,MHz significantly less than some 15-20 per cent. Except for the nucleus, a low thermal fraction is 
implied by steep radio spectra of discussed regions. Even in case of maximum value of 20 per cent, its neglecting leads to an overestimate of magnetic field
strength by no more than 5-7 percent. Therefore, we decided not to take thermal emission into account. 

\begin{table*}
\caption{Parameters used for the estimation of the magnetic field properties and resulting values}
\begin{center}
\begin{tabular}[]{lccccccc}
\hline
\hline
Region & {\it D}\,[kpc]& $\alpha$ & $S_{4.86}$\,[mJy] &$B_{\mathrm{TOT}}$ [$\mu$G]&$E_{\mathrm{B}}$ [$\mathrm{erg}\,\mathrm{cm^{-3}}$] \\
\hline
Core  & 0.65 $\pm$ 0.15 & 0.64 $\pm$ 0.08  & 3.24 $\pm$ 0.16 &  38.8 $\pm$ 1.7 &  60 $\pm$  5  $\times 10^{-12}$ \\
NW    &  5.5 $\pm$ 0.5  & 0.81 $\pm$ 0.07  & 3.59 $\pm$ 0.13 &  12.0 $\pm$ 0.4 & 5.8 $\pm$ 0.5 $\times 10^{-12}$ \\
S     &  5.5 $\pm$ 0.5  & 0.89 $\pm$ 0.06  & 0.57 $\pm$ 0.04 &   9.9 $\pm$ 0.3 & 3.9 $\pm$ 0.3 $\times 10^{-12}$ \\
E     &  5.5 $\pm$ 0.5  & 1.03 $\pm$ 0.19  & 0.08 $\pm$ 0.02 &   8.7 $\pm$ 0.6 & 3.0 $\pm$ 0.4 $\times 10^{-12}$ \\
Ridge &  5.5 $\pm$ 0.5  & 1.01 $\pm$ 0.12  & 0.16 $\pm$ 0.03 &   9.2 $\pm$ 0.6 & 3.3 $\pm$ 0.3 $\times 10^{-12}$ \\
\hline
\end{tabular}
\end{center}
\label{params}
\end{table*}

For the NW part of the ring, we have used a pathlength of 5500\,pc -- derived from the angular size of the emitting region under an 
assumption of a cylindrical symmetry. 
The same size has been also adopted for all the other regions within the disk of NGC\,2445 as well as for the ridge between it and 
NGC\,2444. The estimated strength of the magnetic field in the NW region is therefore $12.0 \pm 0.4\,\mu$G, 
and the energy density is $5.8 \pm 0.5 \times 10^{-12} \mathrm{erg}\,\mathrm{cm^{-3}}$. This means that the magnetic field in this
region is somewhat stronger than the average found in normal, spiral galaxies ($9 \pm 1.3\,\mu$G -- \citealt{niklas}). It is also 
comparable to that found in the shock region between the galaxies forming the Stephan's Quintet ($11.0 \pm 2.2\,\mu$G -- \citealt{SQ}).
Values derived for the other star--forming regions are very similar (see Table~\ref{params}).

Estimates for the intergalactic ridge do not differ much from that for the star--forming regions. As its spectrum is steeper
than that of the star--forming regions (except for the eastern region, which has a comparable $\alpha$), the strength of the magnetic field 
is nearly the same despite lower surface brightness. For an intergalactic structure, value of $\approx 9 \mu$G is a rather high strength. 
There is no clear counterpart to this entity in any other spectral domain. Most possibly, the magnetic field --
enhanced in the star--forming NW region -- is dragged with the intergalactic gas during the tidal interaction with the companion galaxy. 
Such a structure can result from a partially ordered B--field, dragged and stretched during the interaction.
The total magnetic field is, in terms of its strength and energy density, similar to that of the iconic colliding pair of galaxies, the 
Taffies \citep{taffy}. Unfortunately, the setup of the low-frequency observations (simultaneous dual-frequency mode 234/612\,MHz)
did not allow to perform full Stokes observations, and we have not detected polarisation in any of the VLA datasets . Polarisation data is 
necessary to confirm or reject this scenario.

As the core of the collisional ring galaxy exhibits a spectrum that suggests a non-thermal origin of the emission ($\alpha \approx 0.64 \pm 0.08$),
we derived estimates for the magnetic field of the core, too. We used a pathlength of 650\,pc -- a mean of its linear dimensions (see Sect.~\ref{result}).
The resulting field is strong, as it reaches $38.8 \pm 1.7\,\mu$G. Its energy density is equal to $6.0 \pm 0.5 \times 10^{-11} \mathrm{erg}\,\mathrm{cm^{-3}}$.
The core is supposed to undergo starburst activity \citep{multiv}, and for a compact region of efficient electron supply such a 
number is not surprising.

The radio continuum counterpart for the tidal tail has not been detected, but we could obtain the upper limit for the magnetic field strength and
energy in its area. We have assumed that the tail has a cylindrical symmetry with a diameter of 20\,kpc \citep{tail} and a steep 
spectral index of 1.0. Once again, we used the \textsc{bfeld} code, calculating the magnetic field strength using information at
234, 612 and 4860\,MHz. The upper limit for the strength was estimated as $< 4.6 \mu$G. The corresponding limit for the magnetic field 
energy is $< 1.1\times 10^{-12} \mathrm{erg}\,\mathrm{cm^{-3}}$. 
However, these values are of high uncertainty, as the estimates of the tail parameters are rather rough. Assuming the radio-emitting medium to be
more shallow (which is possible, as there are hints of narrowing in certain regions of the tail), the limit for the magnetic field strength 
would rise up -- to approximately 8.5--12\,$\mu$G (in case of the depth of 5 and 1\, kpc, respectively) and its energy density would reach 
2.8 -- 6.3 $\times 10^{-12} \mathrm{erg}\,\mathrm{cm^{-3}}$. This indicates that a relatively strong magnetic field could remain undetected.

\subsection{Age of the structures}
\label{age}

Good frequency coverage allows to estimate the spectral age of the core and the NW region. The amount of time elapsed 
since the last acceleration of the particles in a given structure can be calculated under assumption that the observed steepening of the radio 
spectrum is caused by the synchrotron and/or inverse-Compton processes. We decided to use the \textsc{synage} package \citep{murgia},
which has an implementation of the Jaffe--Perola (\citealt{JP}, JP), Kardashev--Pacholczyk (\citealt{kardashev};~\citealt{pacholczyk}, KP), and
continous injection (\citealt{pacholczyk};~\citealt{myears};~\citealt{carilli}, CI) models of electron energy losses. The JP model assumes that the particles get 
isotropised in the pitch angle with the time-scale of isotropisation much smaller than the radiative lifetime. The KP model assumes that each electron
maintains its original pitch angle. The CI model includes the continuous injection of a power-law distributions of relativistic electrons. The observed
spectrum in this case is the sum of the emission from the various electron populations at different synchrotron ages, ranging from zero to the age of the
source. The flux density at frequencies below the synchrotron break rises with time, since new particles are being added. All these models assume
a constant magnetic field. \textsc{synage} uses the flux values at different wavelengths (spectral energy distribution, SED) to determine the spectral 
index of the injected electron population $\alpha_{\mathrm{inj}}$, and the spectral break frequency $\nu_{\mathrm{break}}$ (frequency above which the 
observed spectrum steepens from the initial one, in GHz). Using the magnetic field strength determined in Sect~\ref{magfield}, 
now expressed in nT, the spectral age can be calculated as:\\
\begin{equation}
\tau = 50.3 \frac{B^{1\slash 2}}{B^2 + B^{2}_{\mathrm{IC}}} \times (\nu_{\mathrm{break}}(1 + z))^{-1\slash 2} \mathrm{[Myr]}
\end{equation}
where $B_{\mathrm{IC}} = 0.338(1 + z)^{2}$ is the CMB magnetic field equivalent \citep{CMB}.
Estimated values of the injection spectral index $\alpha_{\mathrm{inj}}$, break frequency $\nu_{\rm{break}}$, and the spectral age are summarised in Table\,
\ref{fits}.

\begin{figure*}
{\includegraphics{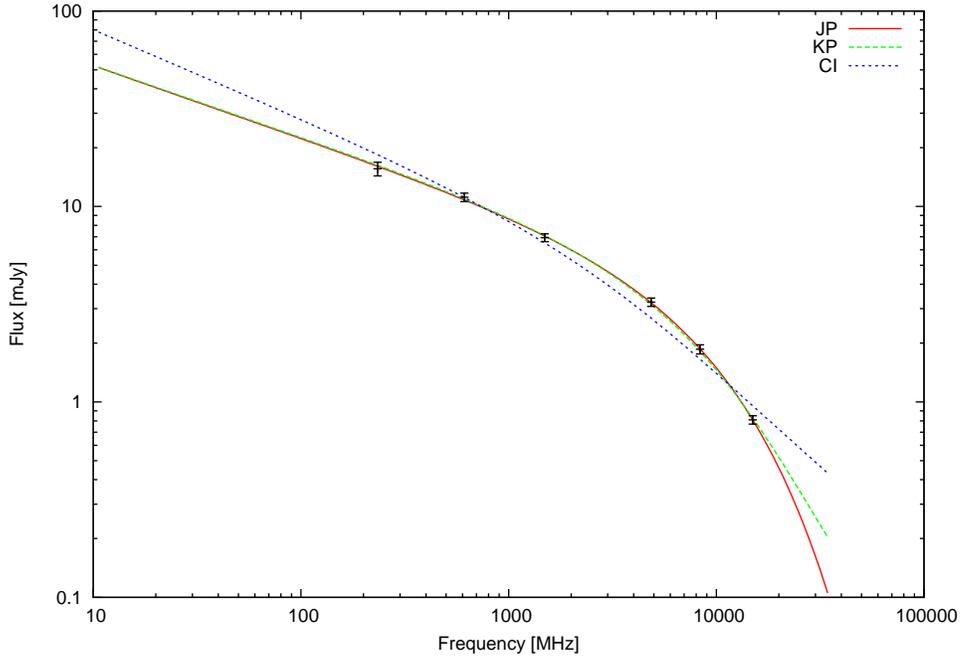}}
\caption{
Three different models of the electron energy losses  models fitted to the SED of the core of NGC\,2445: Jaffe-Perola (JP -- red), Kardashev-Pacholczyk 
(KP -- green) and continous injection (CI -- blue). Details of the fitting scheme and characteristic fit parameters are provided in Sect.~\ref{age}.
}
\label{agecore}
\end{figure*}

Both standard JP and KP models give a good representation of the data, but the fit derived using the JP model is slightly better (judging on the $\chi^{2}$
value -- see Fig.~\ref{agecore}). The fit given by the CI model is worse, as the goodness-of-fit parameter $\chi^{2}$ is of an order of magnitude higher than 
for JP and KP. No matter the chosen model, it is clearly visible that the spectral age of the core is very low, ranging from 1.6 to 5.2\, Myrs. Bearing in mind 
the possible starburst occurrence \citep{multiv} this is a reasonable value; starburst activity would result in an efficient supply of young electrons, 
producing a flat spectrum. The derived spectral index of the injected electron population (significantly below 0.5 for JP and KP, and $0.46\pm 0.05$ for CI) 
indeed suggests a very flat, albeit still reliable spectrum; \citet{SNR} give $\alpha=0.3$ as the flattest spectrum available for the supernova remnants (SNR),
and supernovae are the main sources of the non-thermal emitting electrons in the star-forming galaxies. Moreover, \citet{hummel} lists several galaxies for which the 
injection spectra have $\alpha$ in range of 0.3--0.5, similar to our estimates, both for the core and the NW region. 

\begin{figure*}
{\includegraphics{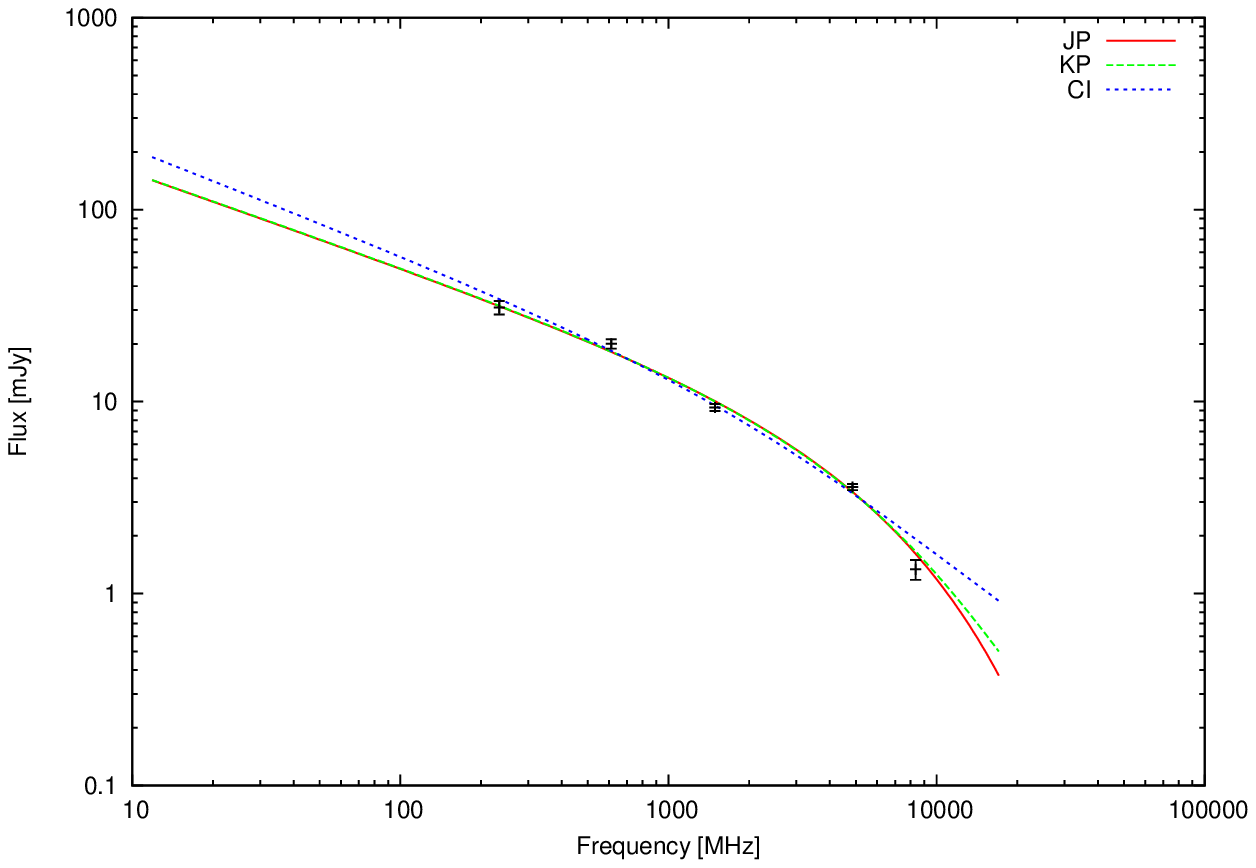}}
\caption{
Three different models of the electron energy losses  models fitted to the SED of the NW region of NGC\,2445: Jaffe-Perola (JP -- red), Kardashev-Pacholczyk 
(KP -- green) and continous injection (CI -- blue). Details of the fitting scheme and characteristic fit parameter are provided in Sect.~\ref{age}.
}
\label{ageshock}
\end{figure*}

The spectral fits for the NW region can be seen in Fig.~\ref{ageshock}. Similarly to the case of the core, the JP and KP fits give better representation than
the CI model. The age estimates range from 9.0 to 39.1\,Myrs. Even in case of the lowest value, the result is higher than the estimates of the age of the 
star-forming knots given by \citet{beirao}, who identifies two very young (2.5 and 3.5\,Myrs) structures in the area coincident with our NW region. This means 
that the high energy electrons associated with the magnetic field are older than those involved in the recent star--forming processes. This is consistent with 
the steep, non-thermal spectrum of this area (Sect.~\ref{spixtext}). Most likely the magnetic field has been enhanced during the propagation of the density wave 
that formed the collisional ring (see \citealt{colrings}), and the recent star--forming activity has weaker effect on its properties (e.g. flattens the spectrum,
but as for it is a non-dominant component, the overall spectrum is still relatively steep). The injection spectral index of around 0.5 is reasonable for an 
extended region of synchrotron radio emission.

Finally, we note that the definite rejection of the CI model on the
basis of available data  may be premature. While for the core the
mentioned fitted spectrum (Fig.~\ref{agecore}) has  generally an overmuch small
curvature, for the NW region the CI model differs from the data  and
other fits only at the high and low ends of the analysed spectrum 
(Fig.~\ref{ageshock}). We keep in mind that the flux measurements at 
low frequencies may be affected by free-free   absorption caused by 
ionised gas. This would decrease the flux at 234\,MHz. At the  high-frequency
end of the radio spectrum, a small $\lambda$/D ratio and a short  
integration time (hence a poor (u,v) plane coverage) of interferometric 
observations may  cause a substantial flux loss of faint extended emission.
With the currently available  data both effects cannot be evaluated quantitatively.
We estimated possible effects of hypothetical free-free absorption and
of loss of  extended structures at lowest and highest frequencies,
respectively upon the results of  CI model fits. We performed an
experiment by increasing arbitrarily the measured flux  densities of the
NW region at extreme frequencies by values needed to improve the 
$\chi^2$ for the CI model. It turns out that achieving similar
goodness-of-fit as for the  JP and KP models for the NW region requires
approximately 10 per cent higher flux at  234\,MHz and 30 per cent
higher flux at 8440\,MHz. We found that the break  frequency, and thus
the age estimate (noted as NW-SIM in Table~\ref{fits}), appeared  to be
similar to those obtained from the original CI fit. In the light of
these estimates and  of the current data status we cannot simply reject
the CI model. Though it implies  (regardless possible flux losses) a
substantially higher limits to the spectral age than JP  and KP spectral
fits, it should be still kept in mind as an acceptable solution for the
NW  region.

\begin{table*}
\caption{\label{fits}Parametres of the spectral fits}
\begin{center}
\begin{tabular}{rrrrrc}
\hline
\hline
Region & Model & $\chi^{2}$ & $\alpha_{\mathrm{inj}}$ & $\nu_{\mathrm{break}} [GHz]$ & Spectral age [Myrs]\\
\hline
Core  & JP &  0.25 & $0.38^{+0.04}_{-0.01}$ & $12.1^{+2.7}_{-1.0}$ &  1.6 --  2.1\\
Core  & KP &  0.48 & $0.38^{+0.03}_{-0.01}$ & $ 6.2^{+1.3}_{-0.5}$ &  2.2 --  2.9\\
Core  & CI & 12.18 & $   0.46\pm 0.05	   $ & $ 2.8^{+1.6}_{-1.0}$ &  2.9 --  5.2\\
NW    & JP &  5.77 & $	  0.49\pm 0.07	   $ & $ 9.3^{+4.8}_{-2.5}$ &  9.0 -- 14.1\\
NW    & KP &  6.34 & $0.49^{+0.08}_{-0.07}$ & $ 5.0^{+3.2}_{-1.4}$ & 11.8 -- 19.5\\
NW    & CI & 10.81 & $0.54^{+0.09}_{-0.10}$ & $ 2.3^{+2.7}_{-1.4}$ & 15.1 -- 39.1\\
NW-SIM& CI &  2.77 & $0.52^{+0.11}_{-0.14}$ & $ 2.8^{+5.0}_{-2.4}$ & 12.1 -- 58.6\\
\hline
\end{tabular}
\end{center}
\end{table*}

\subsection{Morphology of the radio emission distribution}

As mentioned in Sect.~\ref{intro}, the collisional ring of NGC\,2445 is far from being circular/elliptical. So is the radio emission distribution
in this galaxy: its SE side is less prominent than the NW one (not only in the radio regime). The emission is weaker -- not only at 234\,MHz (Fig.~\ref{pband}), but also in better
quality maps at 612 (Fig.~\ref{gband}) and 4860\,MHz (Fig.~\ref{cband}). Star-forming regions in the SE part have also somewhat weaker magnetic fields
and are less extended. Additionally, the angular distance from the core to the ring is lower in the NW part than in the SE part. The asymmetry may result
from an ongoing interaction with NGC\,2444, resulting in the compression of the NW part of the ring, which leads to the amplification of the magnetic field.
Such interaction has been suggested multiple times (e.g. \citealt{multiv}, \citealt{beirao}). 
The distribution of the radio-emitting medium is very similar to that of the molecular emission (\citealt{beirao}),
including lack of the PAH radio emission in the southernmost knot (labeled {\it D} in \citealt{beirao}) and weak emission in the northernmost knot ({\it G} in 
\citealt{beirao}). Both these knots have the lowest far-UV star-forming rates. This suggests that the radio emission deficiency might be result of faster
CR diffusion when compared to the synchrtotron losses. Weaker magnetic field -- especially in the southern part -- allows CR electrons to dffuse regardless of
their energy.

In general, the overall radio distribution seems to confirm the scenario of an off-axis collision and later, on-going interaction with NGC\,2444.
The radio ring follows the distribution of the neutral and ionised gas (\citealt{higdon}, \citealt{beirao}), which is suspected to have formed as a 
result of such collision. The asymmetry of the emission together with the intergalactic ridge provides an
evidence of an interaction with the companion galaxy. However, further details of the interaction history would need examination of the regular magnetic
field properties, i.e. analysis of the polarisation data (like that was done for NGC\,3627 by \citealt{3627}, or the Virgo cluster spirals, by \citealt{virgo}),

\section{Conclusions} 
\label{conclusions}

We observed the tight galaxy pair Arp\,143 with the GMRT at 234 and 612\,MHz. Radio emission maps were analysed together with the archive
VLA data at 1490, 4860, 8440, and 14940\,MHz to study the spectral age of the features seen in the system, morphology of the radio emitting medium 
and the magnetic field strength and energy within the pair. The results obtained are summarised below:\\
-- The radio emission from NGC\,2445 concentrates in the northwestern part of the optical ring, extending to the outer parts of NGC\,2444.
Several distinct radio emitting regions can be identified: the core, three star-forming regions, and an intergalactic ridge.\\
-- The galactic core is a not very compact (0.8 on 0.5\,kpc), synchrotron--emitting source. It possesses a strong magnetic field of 
$ 38.8 \pm 1.7 \mu$G. It has a relatively flat spectrum ($\alpha = 0.64 \pm 0.08$) and the age estimate yields $1.6-5.2$\,Myrs (depending
on the model fitted). This is consistent with its identification as a starburst region.\\
-- The northwestern region of radio emission, coincident with the star--forming knots, possesses magnetic field with strength of $12.0 \pm 0.4 \mu$G
-- somewhat stronger than the typical galactic fields. Its spectrum is typical for an ageing, yet not very old electron population, with a mean 
spectral index of 0.81 $\pm$ 0.07. The spectral age estimate of $9.0-39.1$\,Myrs is higher than that of the star-forming knots given by 
\citet{beirao}, indicating that most of the radio emission is associated with an older electron population, that was injected while a density 
wave was propagating through the intergalactic medium.\\
-- Two other -- southern and eastern -- regions of star-formation have been detected. Their magnetic fields ($ 9.9 \pm 0.3$ and
$ 8.7 \pm 0.6 \mu$G, respectively) are somewhat weaker than that of the northwestern one.\\
-- Distribution of the radio emission from NGC\,2445 shows significant asymmetry, with the northwestern part being more luminous than the 
southeastern one. This supports the scenario of an ongoing interaction with NGC\,2444.\\
-- NGC\,2444 is in general a non-radio-emitting galaxy, apart from the intergalactic radio ridge connecting it with NGC\,2445. The magnetic
field in the ridge is comparable to those found in the star-forming regions. It is also similar to that found in the intergalactic 
bridge of the Taffy galaxies \citep{taffy}.\\
-- There are no signs of radio emission from the {\rm H}{\sc i} tail reported by \citet{tail}. 
Depending on the -- yet unknown -- depth of the radio emitting medium, the upper limit for the magnetic field in the tidal tail is $< 4.5-12\,\mu$G, yielding
$< 1.1-6.3\times 10^{-12} \mathrm{erg}\,\mathrm{cm^{-3}}$ for its energy density.

\section*{acknowledgements}

We wish to thank the anonymous refferee, whose comments and suggestions allowed us to 
significantly improve this article.
We thank the staff of the GMRT that made these observations possible. GMRT
is run by the National Centre for Radio Astrophysics of the Tata Institute
of Fundamental Research.
BNW is indebted to Eric Greisen, NRAO, for his help in preparing the RGB composite
maps using the \textsc{aips}.
This research has been supported
by the scientific grant from the National Science Centre (NCN), dec.
No.\,2011/03/B/ST9/01859.
This research has made use of
the NASA/IPAC Extragalactic Database (NED) which is operated by the Jet Propulsion
Laboratory,
California Institute of Technology, under contract with the National Aeronautics and
Space Administration. 
This research has made use of NASA's Astrophysics Data System.
Funding for the SDSS and SDSS-II has been provided by the Alfred P. Sloan
Foundation, the Participating Institutions, 
the National Science Foundation, the U.S. Department of Energy, the National
Aeronautics and Space Administration, 
the Japanese Monbukagakusho, the Max Planck Society, and the Higher Education
Funding Council for England. 
The SDSS Web Site is http://www.sdss.org/.


\label{lastpage}

\end{document}